\documentclass[10pt, conference]{IEEEtran}
\usepackage{tikz}
\usepackage{truncate} 
\usepackage{filecontents}

\usepackage[colorlinks, citecolor=blue]{hyperref}
 \usepackage{balance}
\usepackage{setspace, amsmath, url, lscape, algorithmic, multirow, pslatex, listings, verbatim, alltt, amsfonts, wrapfig, boxedminipage, color, bookmark,comment}
\usepackage[vlined,linesnumbered,ruled,boxed]{algorithm2e}
\usepackage{array}

\usepackage{subcaption}

\DeclareFontFamily{\encodingdefault}{\ttdefault}{\hyphenchar\font=`\-}
\newcommand\ProcessThreeDashes{\llap{\color{cyan}\mdseries-{-}-}}

\usepackage{paralist}
\usepackage{epsfig}
\let\labelindent\relax
\usepackage[inline]{enumitem}

\usepackage{xcolor}

\usepackage{listings}

\newcommand\YAMLcolonstyle{\color{red}\mdseries}
\newcommand\YAMLkeystyle{\color{black}\bfseries}
\newcommand\YAMLvaluestyle{\color{blue}\mdseries}

\makeatletter

\newcommand\language@yaml{yaml}

\expandafter\expandafter\expandafter\lstdefinelanguage
\expandafter{\language@yaml}
{
  keywords={true,false,null,y,n},
  keywordstyle=\color{darkgray}\bfseries,
  basicstyle=\ttfamily,                                 
  sensitive=false,
  comment=[l]{\#},
  morecomment=[s]{/*}{*/},
  commentstyle=\color{purple}\ttfamily,
  stringstyle=\YAMLvaluestyle\ttfamily,
  moredelim=[l][\color{orange}]{\&},
  moredelim=[l][\color{magenta}]{*},
  moredelim=**[il][\YAMLcolonstyle{:}\YAMLvaluestyle]{:},   
  morestring=[b]',
  morestring=[b]",
  literate =    {---}{{\ProcessThreeDashes}}3
                {>}{{\textcolor{red}\textgreater}}1     
                {|}{{\textcolor{red}\textbar}}1 
                {\ -\ }{{\mdseries\ -\ }}3,
}

\lst@AddToHook{EveryLine}{\ifx\lst@language\language@yaml\YAMLkeystyle\fi}
\makeatother

\DeclareMathDelimiter{(}{\mathopen} {operators}{"28}{largesymbols}{"00}
\DeclareMathDelimiter{)}{\mathclose}{operators}{"29}{largesymbols}{"01}

\newcommand{\qedd}{\nobreak \ifvmode \relax \else
      \ifdim\lastskip<1.5em \hskip-\lastskip
     \hskip1.5em plus0em minus0.5em \fi \nobreak
      \vrule height0.75em width0.5em depth0.25em\fi}

\newcommand{\eg}{{e.g., }}

\newcommand{\agent}{EdgeWeaver agent\xspace}
\newcommand{\agents}{EdgeWeaver agents\xspace}
\newcommand{\name}{EdgeWeaver\xspace}
\newcommand{\namens}{EdgeWeaver}

\newcommand{\comments}[1]{}
\newcommand\hl{\bgroup\markoverwith
  {\textcolor{yellow}{\rule[-.5ex]{2pt}{2.5ex}}}\ULon}

\newcommand{\hai}[1]{{\color{magenta} \textbf{Hai:} #1}}

\begin{document}


\title{\name: Accelerating IoT Application Development Across Edge-Cloud Continuum}

\author{
\IEEEauthorblockN{Pawissanutt Lertpongrujikorn\textsuperscript{1}, Juahn Kwon\textsuperscript{1}, Hai Duc Nguyen\textsuperscript{2}, and Mohsen Amini Salehi\textsuperscript{1}}
\IEEEauthorblockA{\textsuperscript{1}High Performance Cloud Computing (\href{https://hpcclab.org}{HPCC}) Lab, University of North Texas (UNT), USA\\
\textsuperscript{2}Argonne National Laboratory and University of Chicago, USA\\
\{pawissanutt.lertpongrujikorn, Juahn.Kwon, mohsen.aminisalehi\}@unt.edu, hai.nguyen@anl.gov}
\thanks{* These authors contributed equally to this work}
}

\maketitle


\begin{abstract}

The rise of complex, latency-sensitive IoT applications across the Edge–Cloud continuum exposes the limitations of current Function-as-a-Service (FaaS) platforms in seamlessly addressing the complexity, heterogeneity, and intermittent connectivity of Edge-Cloud environments. Developers are left to manage integration and Quality of Service (QoS) enforcement manually, rendering application development complicated and costly. To overcome these limitations, we introduce the \name platform that offers a unified ``object'' abstraction that is seamlessly distributed across the continuum to encapsulate application logic, state, and QoS. \name automates ``class'' deployment across edge and cloud by composing established distributed algorithms (e.g., Raft, CRDTs)---enabling developers to declaratively express QoS (\eg availability and consistency) desires that, in turn, guide internal resource allocation, function placement, and runtime adaptation to fulfill them.
We implement a prototype of \name and evaluate it under diverse settings and using human subjects. Results show that \name boosts development productivity by 31\%, while declaratively enforcing strong consistency and achieving 9 nines availability, 10,000× higher than the current standard, with negligible performance impact.
\end{abstract}

\section{Introduction}\label{sec:intro}

The emergence of the IoT-Edge-Cloud paradigm has transformed the IoT landscape from simple sensor-based systems to complex, data-driven applications that underpin modern smart services across a wide range of domains, such as intelligent transportation \cite{devarajan2024next}, industrial automation \cite{jin2025cloud}, healthcare \cite{alsabah2025comprehensive}, and smart cities \cite{zeng2024sensors}. However, this paradigm shift introduces new challenges, including ensuring desired performance, availability, and consistency across highly heterogeneous and intermittently connected computing tiers, thereby complicating application development and deployment.

\begin{figure}
    \centering
    \includegraphics[width=0.9\linewidth]{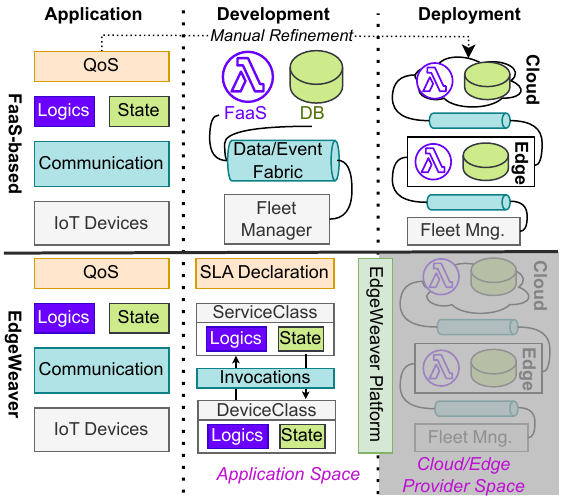}
    \vspace{-2mm}
    \caption{\name vs. FaaS approach to develop and deploy applications across Edge-Cloud continuum. 
    }
    \label{fig:faas-vs-oaas}
    \vspace{-6mm}
\end{figure}

To address these challenges, Serverless Computing, or Function-as-a-Service (FaaS), introduces the function abstraction to hide underlying infrastructure \cite{golec2024cold}. Yet, as shown in the upper part of Figure~\ref{fig:faas-vs-oaas}, the abstraction still required significant efforts. Developers need to integrate state externally (e.g., through databases) and coordinate communication using data/event fabrics such as MQTT and Kafka \cite{kafka}, while considering QoS factors for performance, availability, and consistency \cite{poojara2025scaling}. Furthermore, as existing FaaS platforms are designed for centralized cloud environments, they struggle to operate efficiently in decentralized, failure-prone edge infrastructures due to heterogeneity in the underlying infrastructure \cite{10122638}.


CAP theorem \cite{brewer2012cap} proves the infeasibility of simultaneously guaranteeing multiple QoS metrics (i.e., consistency, availability, and partition tolerance) in such distributed systems. As a result, developers have to make difficult trade-offs via prioritizing certain QoS desires over others (e.g., consistency over availability) based on the application goals and execution conditions \cite{chaudhary2025comprehensive}, which further complicates the development of scalable, resilient IoT systems \cite{russo2023serverless, josbert2024look}. 
Prior efforts have tackled parts of this problem through proposing high-level abstractions \cite{lertpongrujikorn2024streamlining,lertpongrujikorn2024object}, SLA-driven scheduling \cite{nguyen2023storm, nguyen2019real}, and middleware for device fleet management \cite{edgexfoundry}. However, these solutions typically focus on isolated aspects, such as optimizing a single performance metric or supporting specific deployment configurations, without offering a unified, flexible framework for the full application life-cycle. This fragmentation forces developers back into the role of system integrators---manually stitching together disparate tools, thereby increasing development complexity and cost.

To overcome these challenges, we propose \textit{\name}, a novel object-based abstraction to unify and streamline the development and deployment of Edge-Cloud IoT applications.
Inspired by the object-oriented programming (OOP) semantics, as shown in the lower part of Figure~\ref{fig:faas-vs-oaas}, \name establishes the notion of \textit{distributed objects} to encapsulate application logic, state, IoT device interactions, and communication patterns within cohesive class-based abstractions. These \textit{classes} are instantiated and managed as distributed \textit{objects} that span across edge and cloud tiers.
We also introduce a declarative Service Level Agreement (SLA) interface that allows
developers to specify QoS requirements, including consistency, availability, and performance, for objects. \name automatically interprets these SLAs to guide resource allocation, function placement, and runtime adaptation, robustly fulfilling QoS requirements with minimal user intervention.

\name realizes object abstraction through its core component, called \textit{distributed class runtimes}, deployed across Edge-Cloud to hide the underlying infrastructure (\eg compute, storage, and network) complexity. This component enables intuitive, location-transparent function calls within or across distributed objects.
For SLA enforcement, \name provides a unified architecture that automatically composes and configures established distributed mechanisms: (i) Raft consensus and CRDTs~\cite{simic2020crdts} for consistency; (ii) adaptive replication strategies for high availability; and (iii) rate-guarantee abstractions~\cite {nguyen2019real, nguyen2025efficient, andreoli2025multi} for performance. The novelty lies in the framework's ability to automatically select and configure these mechanisms based on high-level SLA specifications. Evaluations demonstrate that \name, as a cohesive platform, can: significantly boost the developer's productivity; maintain reliable QoS under dynamic conditions; and match the efficiency of state-of-the-art systems.
The contributions of this paper are as follows:
\begin{itemize}[leftmargin=*]
    \item Comprehensive Object Abstraction: We introduce a unified distributed object-based abstraction across Edge-Cloud that streamlines the development and deployment of IoT applications. Our evaluation shows this abstraction reduces implementation by up to 44.5\% and configuration code by up to $10\times$ fewer and, through a human study, demonstrates 31\% development time reduction compared to conventional FaaS approaches (\S\ref{sec:approach}).
    \item Automated Algorithm Composition via Declarative SLAs: We developed a declarative SLA mechanism that enables automated, self-adaptive QoS enforcement. This allows applications to achieve strong consistency and up to 9 nines availability—$10{,}000\times$ more reliable than the current standard at only $3\times$ the cost (\S~\ref{sec:realization}).
    \item Systematic evaluation: Real-world experiments show that \name scales throughput by up to $70\times$ while automatically adapting to dynamic environments and maintaining SLA compliance with minimal manual effort. (\S~\ref{sec:evaluation}). 
\end{itemize}


\section{Problem Statement}
\label{sec:background}

\begin{figure} [ht]
    \centering
    \includegraphics[width=1\linewidth]{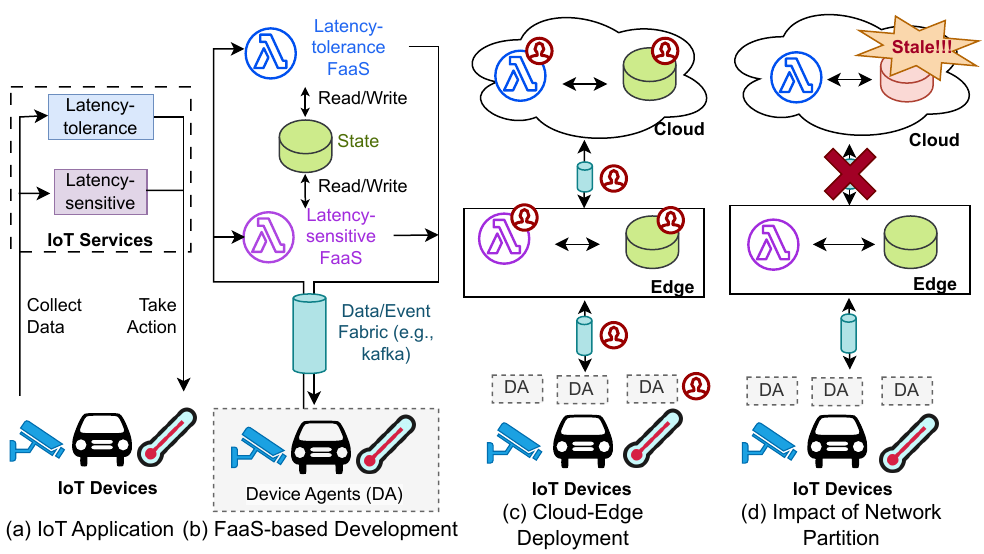}
    \vspace{-5mm}
    \caption{FaaS-based development and deployment challenges}
    \label{fig:faas-iot}
    \vspace{-4.5mm}
\end{figure}

As illustrated in Figure \ref{fig:faas-iot}a, modern IoT applications often rely on devices that continuously collect data from their surroundings. For instance, traffic detection systems constantly gather video from street cameras~\cite{sharma2025video}. These data streams are then processed by IoT services, which may be latency-sensitive (e.g., speed violation detection \cite{daraghmi2022iot})  or latency-tolerant (e.g., analyzing traffic patterns to optimize signal timing \cite{ michael2025traffic}). Modern FaaS deployments adopt the Edge–Cloud paradigm, where resource-constrained edge servers near data sources handle latency-critical functions, while the others are offloaded to the cloud (Figure \ref{fig:faas-iot}d) \cite{russo2023serverless, russo2025towards}. Although this hybrid architecture enhances responsiveness and resource efficiency, it also introduces significant design and operational challenges.

\subsection{Complexity of Application Development}
The ``function'' abstraction in FaaS focuses solely on application logic, leaving data management and communication to developers. This results in fragmented implementations requiring complex interactions among FaaS invocations, databases, and orchestrators \cite{lertpongrujikorn2024streamlining}. Developers must navigate multiple interfaces, complicating application design. Furthermore, the stateless nature of FaaS \textit{lacks built-in support for long-lived functions}, which are crucial for IoT scenarios involving frequent or continuous data processing. Consequently, developers must manually coordinate intricate dataflows across many short-lived functions and shared-state services, making robust, end-to-end architectures difficult to maintain \cite{patsch2024make, lertpongrujikorn2024streamlining}.


In parallel, \textit{heterogeneity} introduces additional challenges in managing and deploying IoT applications across the Edge–Cloud continuum. First, due to resource limitations, FaaS abstractions on the edge (e.g., using k3s) differ from their cloud counterparts (e.g., using k8s), increasing deployment complexity \cite{dritsas2025survey}. Second, large-scale IoT solutions span diverse technologies, application models, and communication protocols \cite{noaman2022challenges, dave2024data}. Although IoT middlewares provide abstraction layers for easier administrations \cite{miquel2025middleware}, these layers add management overhead and require service providers to handle fleet-wide configurations, deployments, and updates.

As a result, expanding IoT solutions across Edge-Cloud requires a significant development effort, especially at scale. As shown in Figure \ref{fig:faas-iot}c, even with a single edge, a developer must implement and manage numerous components (marked with red head icons). With additional edge sites, this complexity escalates, as each node may differ in software stacks, capacity, network conditions, and policies. Coordinating such heterogeneous environments remains a major challenge \cite{carvalho2021edge, aslanpour2024load},\textit{ highlighting the need for new abstractions to improve integration, scalability, and variability in FaaS-based IoT systems}.

\subsection{No One-size-fit-all Solutions}
\label{sec:no-one-size-fit-all}


The Edge-Cloud continuum often suffers from \textit{intermittent connectivity} between tiers, caused by network congestion, physical obstructions, power limitations, or fluctuating bandwidth \cite{esteves2024long}.
As depicted in Figure \ref{fig:faas-iot}d, lost or delayed connections (i.e., \textit{network partition}) can disrupt data flow, causing communication delays or losses. 
This undermines consistency, as the cloud's view of the application's state becomes stale while the edge node continues processing data locally, leading to serious consequences: unsynchronized data on a production line may result in incorrect machine behavior, and outdated patient information could lead to delayed or even dangerous clinical decisions \cite{abbas2024iomt}. Waiting for cloud confirmation ensures consistency but reduce availability, potentially delaying urgent actions like dispatching an ambulance after an accident \cite{nguyen2019real}.


In fact, the \textit{CAP theorem} \cite{brewer2012cap, lee2021quantifying} formally shows that simultaneously guaranteeing \textit{consistency} and \textit{availability} under network partitions is impossible. Consequently, Edge–Cloud IoT deployments must carefully adapting their design to balance trade-offs among consistency, availability, and performance for their application’s specific QoS needs. This challenge is compounded by the heterogeneity of IoT environments, where devices, infrastructure, and communication protocols differ widely. Ensuring both timely responses and accurate data across such diverse and intermittently connected systems remains a core difficulty in Edge–Cloud IoT design.

\section{Related Studies}
\label{sec:related-work}

\subsection{Managing Edge-Cloud Complexity}

\noindent\textbf{Streamlining Edge-Cloud deployments.}
Contemporary Edge–Cloud solutions let applications express QoS (e.g., latency, availability, and cost) as SLAs to guide automatic FaaS placement and scheduling, shielding developers from low-level system heterogeneity \cite{goudarzi2022scheduling, nguyen2023storm}. The approach is supported by extensive research, often by modeling it as an optimization problem to minimize cost \cite{pusztai2025chunkfunc, russo2024qos, yao2023performance, baresi2024neptune,  liu2023faaslight} or energy consumption \cite{aslanpour2024faashouse, vahabi2023energy} under SLA and resource constraints.
Beyond QoS-based scheduling, fleet management frameworks address the operational complexity of large, heterogeneous IoT deployments. Middleware platforms such as Eclipse IoT projects \cite{eclipseiot-ditto} and EdgeX Foundry \cite{edgexfoundry} provide unified interfaces, protocol translation \cite{dafare2023lora,beniwal2022systematic}, and centralized registries (e.g., AWS IoT Device Management, Azure IoT Hub) for scalable onboarding, security, and monitoring. Automated tooling \cite{santos2023automated,alfonso2023model} ensures consistent configuration and deployment, while policy-driven orchestration \cite{lea2025internet} applies tiered control to maintain reliability and scalability across diverse device fleets.


\noindent\textbf{Unified compute-data abstraction.}
Many serverless platforms integrate functions and state into cohesive units, such as actors \cite{spenger2024survey} or proclets \cite{ruan2023nu}, to streamline stateful function deployment. Azure Entity Functions \cite{azure_enfunc}, inspired by the virtual actor paradigm (e.g., Orleans \cite{bykov2011orleans}), exemplify this design by using conflict-free replicated data types (CRDTs) \cite{shapiro2011conflict} for consistent state replication. Similarly, OaaS  \cite{lertpongrujikorn2023object,lertpongrujikorn2024object,lertpongrujikorn2024streamlining} and Nubes \cite{marek2023nubes} apply object-oriented principles to encapsulate functions, state, and non-functional requirements into unified, cloud-native deployments.

\color{black}

\subsection{Handling Intermittent Connectivity}

Due to the CAP theorem's constraints, handling intermittent connectivity across the Edge-Cloud continuum inevitably requires a trade-off between availability and consistency. As a result, developers must make application-specific decisions based on architectural needs and QoS priorities.

\vspace{1mm}
\noindent\textbf{Providing high availability.}
For applications prioritizing availability, weaker consistency models like causal or eventual consistency are often adopted \cite{ bailis2013eventual}. These models use CRDTs \cite{toumlilt2021highly}, versioning, gossip protocols \cite{ casadei2019self}, and session guarantees \cite{trivedi2020sharing} to handle data divergence and maintain service during network outages. Several studies let applications specify data requires strong consistency vs those can tolerate staleness. The Edge–Cloud orchestration then adjusts function placement and replication accordingly \cite{lfedgeeve,azurecosmos}.

\noindent\textbf{Enforcing strong consistency.}
When strong consistency is crucial, developers often replicate deployments across multiple Edge-Cloud regions and use consensus protocols like Paxos \cite{lamport2001paxos} or Raft \cite{ongaro2014search} to ensure updates are agreed upon by a majority of replicas before being committed. While ensuring correctness, doing so reduces availability during network failures, as writes are paused until connectivity and synchronization are restored. To mitigate availability concerns, Google Cloud Spanner \cite{google-cloud-spanner} leverages globally synchronized clocks with bounded error to offer strong consistency and high availability. CockroachDB \cite{taft2020cockroachdb} uses multi-version concurrency control and consensus to maintain consistency, deployment flexibility, and ease of migration.

\begin{figure*} [t]
    \centering
    \includegraphics[width=0.80\textwidth]{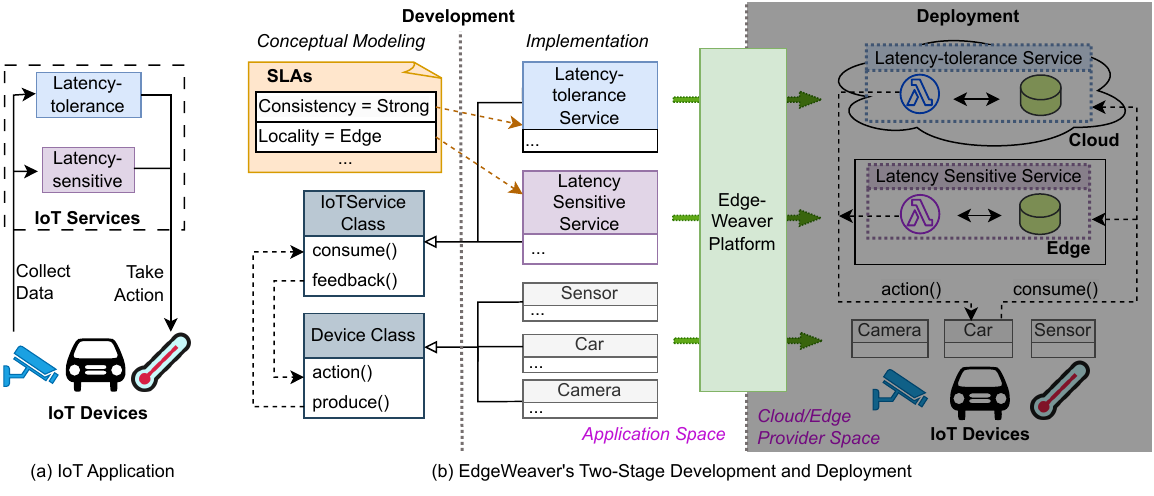}
    \vspace{-2.5mm}
    \caption{Overview of the \name paradigm to resolve Edge-Cloud challenges for IoT applications}
    \label{fig:oaas-iot}
    \vspace{-5mm}
\end{figure*}

\subsection{Limitations}

\noindent\textbf{Lack of a comprehensive solution}.
Most efforts to simplify application deployment address \textit{ only isolated aspects} of the challenge rather than the \textit{entire application lifecycle}. For example, OaaS \cite{lertpongrujikorn2024streamlining} focuses primarily on cloud environments, where SLA enforcement often targets a single metric (e.g., throughput \cite{nguyen2019real}) while overlooking other key factors such as availability and consistency. As a result, developers still expend significant effort integrating these \textit{fragmented solutions} into cohesive and robust Edge–Cloud IoT systems \cite{9474932, rahmani2025optimizing}.

\vspace{1mm}
\noindent

\vspace{1mm}
\noindent
\textbf{Lack of Unified SLA Support.}
Existing frameworks often treat QoS on a fragmented basis. They either offer very limited SLA support (e.g., Ekko \cite{ekko} provides no SLA/QoS guarantees) or implement fundamentally narrow SLA models (e.g., TEMPOS \cite{tempos} focuses exclusively on real-time execution guarantees). Consequently, there remains a gap for a unified SLA framework that concurrently spans performance, availability, and consistency.

\vspace{1mm}
\noindent\textbf{Inflexibility}.
Current solutions are tightly bound to \textit{specific} application requirements and Edge-Cloud configurations, limiting the ability to reconfigure, expand, or migrate Edge-Cloud IoT services. Many commercial offerings are confined to a single public cloud or edge provider, making it difficult to construct hybrid deployments that span private infrastructures or multiple vendors \cite{firdaus2024exploring}. Additionally, due to the lack of a \textit{unifying framework}, even minor updates, such as changing a service configuration or upgrading an SLA, can result in \textit{unpredictable ripple effects}, requiring extensive manual effort to ensure system correctness. This inherent rigidity hampers robust maintenance and scaling, thereby significantly increasing the cost of evolving Edge-Cloud IoT solutions.



\section{\name: Abstraction for the Continuum}
\label{sec:approach}

To overcome the limitations of existing approaches in addressing the complexity and connectivity challenges of IoT deployment across the Edge–Cloud continuum, we present the \name platform. \name provides a \textit{unified abstraction} that \textit{decouples} the application from the underlying infrastructure, reducing the effort required to develop, adapt, and maintain IoT services in heterogeneous environments.

\subsection{Comprehensive Object Abstraction}

Inspired by unified compute-data abstractions in the cloud \cite{lertpongrujikorn2024streamlining, azure_entity, copik2024process}, \name adopts object-oriented programming (OOP) principles to provide a comprehensive abstraction for IoT applications through the notion of ``object'' abstraction. In this abstraction, applications consist of \textit{distributed objects}, each of which contains \underline{\textit{attributes}} representing IoT data or state.
\underline{\textit{Functions (or methods)}} encapsulate application logic, implemented as serverless functions.
\underline{\textit{Communication}} occurs through method invocations between objects, abstracting control and data flow across edge and cloud. \underline{\textit{QoS constraints}} (e.g., performance, availability, consistency) are declared as SLAs and enforced transparently by the platform. The abstraction also includes deployment constraints (e.g., locality) to relieve the burden of low-level configuration (e.g., choosing where to deploy an object) from developers (see Table \ref{tab:non-functional-requirements}).

This unified abstraction provides a comprehensive and intuitive view of the heterogeneous IoT devices and services while hiding implementation details, relieving developers from low-level system management. 
By establishing key OOP application-building features such as abstract classes, inheritance, and polymorphism across the edge–cloud continuum, \name promotes software reuse and modularity. Developers can define core functionalities as base classes and extend them to meet specific use cases, enhancing flexibility and maintainability. Furthermore, \name leverages OOP access modifiers and object composition features to organize application structure and encapsulate application flow through function calls. Crucially, \name abstracts away infrastructure dependencies, requiring no specific configurations or underlying technologies. As a result, applications remain fully decoupled from deployment environments and can be effortlessly deployed across diverse setups.

\vspace{-2mm}
\subsection{Declarative SLA-Driven Deployment}
\name abstracts deployment complexity through SLA specifications, allowing developers to define non-functional requirements, such as performance, availability, and consistency, at the class level or for individual attributes and methods. These SLAs guide the platform's automated deployment process, ensuring that applications meet their QoS targets without manual intervention. By structuring application logic and state as objects, \name has a unified view to efficiently allocate resources, proactively distributing functions and associated data together to minimize latency and data transfer overhead. This SLA-driven resource management not only improves runtime efficiency but also adapts to dynamic conditions and infrastructure heterogeneity, enabling seamless and reliable execution across Edge-Cloud continuum.

\subsection{Application Development and Deployment}

As illustrated in Figure \ref{fig:oaas-iot}, the \name establishes a novel two-stage abstraction for developing and deploying IoT applications across Edge-Cloud continuum:

    \underline{\textit{Conceptual Modeling}}: Developers begin by modeling essential components and workflows using \name’s unified abstraction. For example, heterogeneous IoT devices are modeled by a \texttt{Device} class, defining their basic operations (e.g., produce data and take action) with other IoT services, which are also modeled as separate classes. This high-level blueprint abstracts away the complexities of the underlying infrastructure, offering a clear and unified design framework.
    
    \underline{\textit{Implementation}}: Developers extend these conceptual classes to capture the specifics of actual IoT devices and services, associating them with SLAs. The enriched class definitions are then submitted to the \name platform. The platform extracts the embedded logic, state, and communication patterns to instantiate concrete Edge-Cloud components (e.g., FaaS functions, databases, and event pipelines) for deployment.
\textit{Deployment is fully automated} within the provider ecosystem, with the declared SLAs driving each component placement, scheduling, and management. For example, as shown in Figure \ref{fig:oaas-iot}b, the latency-sensitive service is implemented with a ``\texttt{locality=Edge}'' SLA. During deployment, \name places its corresponding FaaS functions and state at the edge nodes to reduce data access and device communication latency. Conversely, a latency-tolerant service that requires strong consistency (e.g., linearization) is deployed on the cloud, where robust infrastructure can meet its SLA demands. With this SLA-driven deployment, \name ensures diverse application QoS requirements are met across Edge-Cloud continuum without any tuning effort from the developers, thereby, mitigating their deployment effort.   

In sum,  \name overcomes the limitations of existing approaches by providing a unified framework that simplifies IoT application development and deployment by abstracting both functional and non-functional aspects, ensuring seamless operation across intermittent Edge–Cloud environments.

\begin{figure}[t]
    \centering    
    \includegraphics[width=0.95\linewidth]{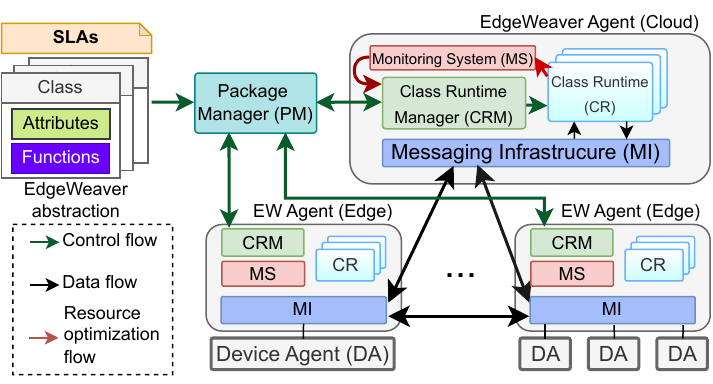}
    \vspace{-2mm}
    \caption{\name Architecture}
    \label{fig:oaas_arch}
    \vspace{-5mm}
\end{figure}

\section{\name Realization}
\label{sec:realization}


\subsection{Design Goals and Architecture}

We design a \name platform, following the architecture shown in Figure~\ref{fig:oaas_arch}, to meet three key requirements:

\vspace{1mm}
\noindent\textbf{a. Comprehensiveness.} To reduce development complexity and provide a unified view of IoT applications, we leverage object-oriented programming (OOP) concepts to define abstractions that capture both functional and non-functional aspects of application logic without exposing developers to underlying infrastructure details. We implement the \textit{Object Abstraction}, which provides APIs and tools for modeling applications as classes, along with SLAs to specify requirements for performance, availability, and consistency. Additionally, we introduce lightweight \textit{Device Agents} (DA) that run on IoT devices, exposing platform-specific APIs to integrate these devices into the \name environment. Together, these components create a unified abstraction layer that supports full lifecycle of application design and development.

\vspace{1mm}
\noindent\textbf{b. Adaptability.} To handle the dynamics of execution environments (e.g, network failures), \name enables automated, SLA-driven adaptation. For that purpose, each tier across Edge–Cloud hosts an \textit{\agent}, which includes a \textit{Monitoring System} (MS in Figure~\ref{fig:oaas_arch}) that collects SLA-related metrics and a \textit{Class Runtime Manager (CRM)} that adjusts deployments in real-time. This forms a localized control loop that continuously adapts to workload fluctuations, resource availability, and network conditions without manual tuning.

\vspace{1mm}
\noindent\textbf{c. Applicability.} 
To ensure broad adoption and scalable operation across heterogeneous infrastructures, \name uses a modular architecture. Objects are deployed via \textit{Class Runtimes (CR)}, configured by the Class Runtime Manager (CRM) according to both SLA specifications and capabilities of the hosting datacenter. Each \agent includes a \textit{Messaging Infrastructure (MI)} that abstracts inter-object communication into a topic-based protocol-agnostic model. At the global level, \textit{Package Manager} coordinates deployment and synchronization across tiers, enabling platform-wide consistency and scalable, hybrid Edge-Cloud deployments.

 

By fulfilling these design requirements, the \name architecture delivers its original vision: simplifying application development; enabling flexible and automated deployment; and supporting robust, SLA-compliant execution in heterogeneous and intermittently connected environments. In the remainder of this section, we describe how the architectural components interact to support end-to-end IoT application conceptualization, development, and deployment---demonstrating how \name meets its design goals in practice.


\begin{table*}[ht]
    \centering
    \small
    \begin{tabular}{|l|p{3.2cm}|l|p{5.1cm}|p{5.1cm}|}
        \hline
        \textbf{SLA} & \textbf{Value Type} & \textbf{Unit} & \textbf{Definition} & \textbf{Algorithmic Mechanism}\\ \hline \hline
        Consistency 
        & Strong & N/A & Read always reflects the latest write. & Raft Consensus: Leader-based replication ensures linearizability. \\
        & Bounded Staleness ($\Delta$) & sec & Read can lag behind the latest write, but only within $\Delta$ seconds. & Merkle Trees + CRDTs: Efficient state synchronization and conflict resolution within time bounds. \\
        & Read your Write (RYW) & N/A & Ensure a client's next read includes its most recent write. & Session Guarantees: Client-centric routing to local replicas. \\ \hline \hline
        Availability & Real & \% & The percentage of time an object/function must be available for service. & Adaptive Replication: Replica count calculated via probabilistic failure models~\cite{meroufel2013managing}. \\ \hline \hline
        Throughput & Integer & RPS & Minimum number of invocations guaranteed to be executed per second. & Resource Pre-warming: Predictive auto-scaling based on real-time Serverless~\cite{nguyen2019real}. \\ \hline
        Locality & Preferred datacenters & N/A & Preferred data centers that will be used for deployment. & SLA-driven Placement: Constraint solving to place data/compute near sources. \\ \hline
    \end{tabular}
    \caption{SLAs Supported by \name and their Corresponding Distributed Mechanisms}
    \label{tab:non-functional-requirements}
    \vspace{-3mm}
\end{table*}

\begin{table*}[ht]
    \centering
    \small
    \begin{tabular}
    {|p{2cm}|l|p{11cm}|}
        \hline
        \textbf{Categories} & \textbf{API} & \textbf{Explanation} \\ \hline \hline
        \multirow{3}{2.5cm}{Object APIs}
            & \texttt{CLASS.create()} & Create a new object of class \texttt{CLASS} and return its ID. \\ \cline{2-3}
            & \texttt{CLASS.get(ID)} & Retrieve an object of class \texttt{CLASS} by ID. \\ \cline{2-3}
            & \texttt{CLASS.delete(ID)} & Delete an object of class \texttt{CLASS} by ID. \\ \cline{2-3} \hline
        \multirow{2}{2.5cm}{Attribute APIs}
            & \texttt{commit(obj, attr)} & Write local changes of \texttt{attr} in \texttt{obj} to storage. \\ \cline{2-3}
            & \texttt{refresh(obj, attr)} & Read the latest value of \texttt{attr} in \texttt{obj} from storage. \\ \cline{2-3} \hline
        \multirow{2}{2.5cm}{Function APIs}
            & \texttt{trigger(func, src, e)} 
                & Trigger \texttt{func} when event \texttt{e} occurs on \texttt{src}. Events: \texttt{OnComplete} or \texttt{OnFailure} if \texttt{src} is a function; \texttt{OnCreate}, \texttt{OnUpdate}, or \texttt{OnDelete} if \texttt{src} is an attribute. \\  \cline{2-3}
            & \texttt{suppress(func, src, e)} & Disable trigger on \texttt{func} from \texttt{src} on event \texttt{e}. \\ \hline
    \end{tabular}
    \caption{\name's API}
    \label{tab:oaas-iot-api}
    \vspace{-7mm}
\end{table*}

\begin{figure} [t]
    \centering
    \includegraphics[width=0.9\linewidth]{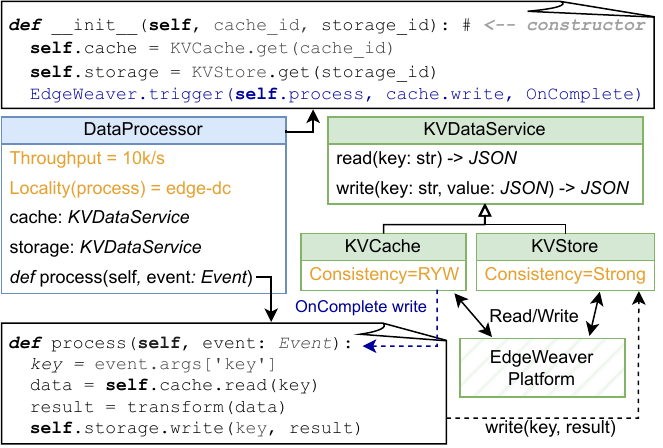}
    \vspace{-2mm}
    \caption{Modeling and implementing a simple IoT processing service with \name}
    \label{fig:oaas-abstraction-example}
    \vspace{-5mm}
\end{figure}

\subsection{Object Abstraction Realization}
\label{sec:realization-abstraction}

At its core, an \name application is structured around classes that define the blueprint for independently executable objects. Each class encapsulates attributes (representing state or data) and methods (implemented as serverless functions), following object-oriented programming principles. Upon deployment, \name automatically instantiates and manages these objects using distributed \textit{Class Runtimes}, as described earlier. The abstraction also supports inheritance and polymorphism, allowing developers to create extensible, reusable components, promoting modular design and reducing code duplication.
EdgeWeaver allows developers to associate SLAs with classes, methods, or attributes to specify QoS requirements, including locality, throughput, availability, and consistency (see Table \ref{tab:non-functional-requirements}). \name also provides a high-level API (Table~\ref{tab:oaas-iot-api}) for inter-object communication, event-driven triggers, and system interactions, all without handling low-level details like networking protocols or deployment scripts.

Figure \ref{fig:oaas-abstraction-example} illustrates how the abstraction supports comprehensive, infrastructure-independent application development. This example has an IoT data processing service, implemented by the \texttt{DataProcessor} class that consumes data from a short-term cache, processes it, and writes it to a long-term store. Since both expose a key-value interface, developers define a base class (\texttt{KVDataService}) and extend it into \texttt{KVCache} and \texttt{KVStore} through inheritance. 
\name manages all object instantiations and data handling internally. Developers simply attach SLA annotations, for example, \texttt{Consistency=Strong} on \texttt{KVStore} to ensure strict linearizability without manually configuring consensus protocols (e.g., Raft \cite{ongaro2014search}). The \texttt{DataProcessor} interacts with these services via class attributes, which are automatically injected by the platform at object creation. Developers can also register event-driven triggers (e.g., executing a function upon adding data to the cache). Furthermore, performance requirements can be specified either globally (e.g., \texttt{throughput=10k/s}) or per method (e.g., \texttt{Locality(process)=edge-dc}). All are managed entirely by \name; without any custom orchestration, messaging setup, or deployment scripting.

\begin{figure}
    \centering
    \includegraphics[width=0.8\linewidth]{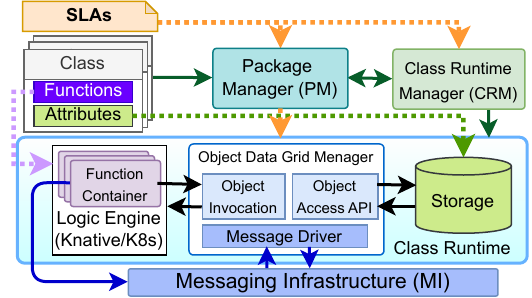}
    \vspace{-2mm}
    \caption{Class Deployment with Class Runtime}
    \label{fig:oaas_cr}
    \vspace{-4.5mm}
\end{figure}

\begin{figure}
    \centering
    \includegraphics[width=0.95\linewidth]{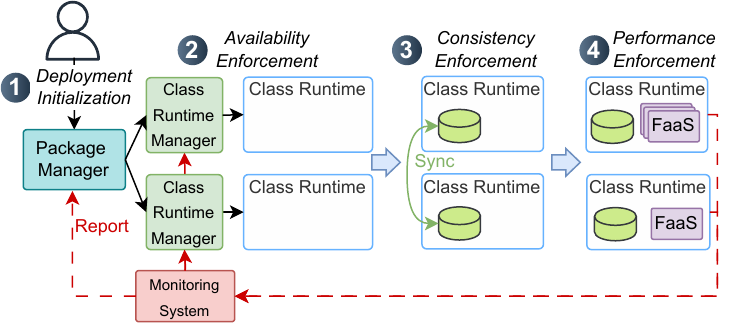}
    \caption{Enforcing SLAs through class deployment}
    \label{fig:oaas-object-slas}
    \vspace{-5mm}
\end{figure}

\subsection{Class Deployment}
\label{sec:realization-object}

Upon development completion, developers submit their applications to \name as a collection of class definitions annotated with SLAs.
Figure~\ref{fig:oaas_cr} illustrates how \name transforms this submission into concrete deployments.
First, the deployment package, including class definitions and SLAs, is submitted to the Package Manager. Based on the list of data centers the application is authorized to access, the Package Manager forwards the relevant class definitions to the corresponding \agents operating in those target data centers. Each destination \agent invokes its Class Runtime Manager to process the submission and instantiate a dedicated Class Runtime for each class. These class Runtimes manage the lifecycle of all object instances associated with the class and enforce their SLA during their lifetime.


Each Class Runtime is composed of three key components: (i) \textit{Logic Engine} that executes class methods; (ii) \textit{Storage System} that instantiates appropriate storage backends for managing object attributes based on their data types and consistency requirements; and (iii) \textit{Object Data Grid Manager (ODGM)} that orchestrates invocations and data access using modules for invocation routing, data consistency, and cross-datacenter communication (via Zenoh~\cite{liang2023performance}).




\subsection{SLA Enforcement}
\label{sec:realization-sla}

Once a class is successfully submitted and deployed, developers can begin instantiating objects from it. These objects represent running instances of the application logic, and their deployment must comply with the SLAs specified in the original class definition.
As summarized in Table~\ref{tab:non-functional-requirements}, \name enforces these diverse SLA targets by automatically composing and configuring appropriate distributed algorithms. Figure~\ref{fig:oaas-object-slas} shows how \name enforces SLAs during deployment and execution, which we detail in the following subsections. These three SLA dimensions operate synergistically: the Package Manager first determines the number and placement of replicas to ensure \textit{availability}; the ODGM then binds these distributed replicas through consensus or synchronization protocols to establish \textit{consistency}; finally, the Class Runtime Manager (CRM) pre-allocates and co-locates compute resources around these storage replicas to guarantee \textit{performance}.

\subsubsection{Availability Enforcement}

At class deployment time, the Package Manager selects appropriate tiers (a.k.a. datacenters) to host Class Runtimes, guided by the availability SLA.
It estimates the failure probability of each data center using metrics (e.g., uptime, network reliability) collected from the Monitoring System. Based on these estimates and the desired availability target, it calculates the required replication factor using the Meroufel and Belalem method~\cite{meroufel2013managing}. The Package Manager then uses the replication factor to select the necessary number of data centers that satisfy developer-specified constraints (e.g., locality). If no constraints are given, it defaults to a round-robin strategy for load balancing. Class Runtimes are then deployed across these selected sites to host object replicas and ensure SLA-compliant availability.

\subsubsection{Consistency Enforcement}
\label{sec:realization-sla:consistency}

When Class Runtimes are deployed, they provision storage and coordinate with each other to enforce the consistency SLA.

\noindent\textbf{Strong Consistency.}
For applications requiring linearizability, \name employs the Raft consensus algorithm~\cite{ongaro2014search}. Raft operates by electing a leader among replicas to manage the replication log. All write requests are routed to the leader, which replicates the entry to followers. A write is committed only after a majority acknowledge it, ensuring the system can survive minority failures without data loss. \name integrates Raft directly into the ODGM’s \textit{Object Access API}, ensuring that all replicas agree on the latest object state before processing reads or writes.

\noindent\textbf{Bounded-Staleness.}
For applications tolerating temporary lag, \name allows stale reads within a time-bound $\Delta$. To manage this, ODGM employs Merkle Search Trees (MST)~\cite{auvolat2019merkle}, a data structure that hashes the storage content hierarchy. Replicas periodically exchange root hashes; if they differ, they traverse the tree to identify divergent keys efficiently, minimizing bandwidth usage during synchronization.
To resolve conflicts during these merges, \name uses Conflict-free Replicated Data Types (CRDTs) \cite{simic2020crdts}, ensuring that state converges mathematically without manual intervention. Read/write access is blocked only if network partitions exceed the allowed staleness window $\Delta$.

\noindent\textbf{Read-Your-Write (RYW).}
This model allows reads to see a recent write from the same client, even under network partition. Class Runtime enforces this by routing reads and writes through the object access API to the same local storage replica, effectively providing ``session guarantees'' without requiring global consensus.
\subsubsection{Performance Enforcement}
Each class’s methods are deployed by the Logic Engine into isolated containers. If a throughput SLA is specified, containers are pre-warmed with sufficient compute resources, calculated using techniques from real-time Serverless~\cite{nguyen2019real}, to meet the required invocation rate. If a locality SLA is present, the preferred data center must reserve enough resources to meet both throughput and co-location requirements. The Class Runtime ensures that containers are deployed on the same machine as the object’s storage to reduce access latency and pre-warms containers to avoid cold starts. When multiple replicas exist and locality is not a constraint, resource allocation is balanced across them, proportional to their resource availability.
\subsubsection{SLA-compliance Execution}
After deployment, users interact with objects via the API provided in Table~\ref{tab:oaas-iot-api}. API calls are routed through the \textit{Messaging Infrastructure}, which transparently directs each request to the appropriate ODGM instance based on the object ID embedded in the calls.
Upon receiving a request, the ODGM’s \textit{Object Invocation} module triggers the corresponding function on the local Logic engine. If the function call targets a remote object, the ODGM leverages the \textit{Message Driver} to relay the request to the corresponding location. This mechanism enables seamless, location-transparent invocation across the Edge–Cloud continuum.
When a function needs to access object attributes, the attribute ID is passed through the \textit{Messaging Infrastructure} to the \textit{Object Access API} of the relevant ODGM. Before any data operation is executed, the ODGM enforces consistency guarantees by running the necessary replication and consistency protocols (see \S\ref{sec:realization-sla:consistency}), to enforce SLA-compliant execution.
\subsubsection{SLA Monitoring and Lifecycle Management}
Monitoring System continuously collects SLA-related metrics (from Class Runtimes) that are reported to the Package Manager and the Class Runtime Manager. Upon SLA violation or runtime failure, \name automatically initiates corrective actions, such as reallocating resources or instantiating new runtimes to maintain SLA compliance  for all objects during their lifecycle.


\vspace{-1mm}
\subsection{Implementation}
\label{sec:implementation}
We implemented a \name prototype utilizing widely adopted open-source technologies:


\noindent\textbf{Control Plane}: The Package Manager and Class Runtime Manager are primarily implemented in Java. We also provide the SDK in Python for class implementation and definition.

\noindent\textbf{Container Orchestrator}: The platform is built on Kubernetes (for cloud) and K3s (for edge), leveraging Knative for serverless function orchestration.

\noindent\textbf{Messaging Infrastructure}: Inter-component communication uses Zenoh~\cite{corsaro2023zenoh}. By design, Zenoh abstracts away physical placement and network routing across the Edge-Cloud continuum, enabling seamless cross-datacenter message passing.

\noindent\textbf{Class Runtimes}: Integrating computation and state, Class Runtimes leverage Knative to orchestrate serverless function containers (Logic Engine). The ODGM, implemented in Rust, acts as a core translation layer that subscribes to Zenoh topics and converts them into local HTTP calls for Knative, bridging the gap without modifying Knative's core. For object state (Storage), we integrate OpenRaft~\cite{openraft} for strong consistency, alongside custom Merkle Search Trees and CRDT wrappers for Bounded-Staleness and RYW. By routing all state synchronization payloads through Zenoh, these algorithms operate seamlessly across the edge-cloud without custom networking logic.

\noindent\textbf{Source code:} including the runtime, SDKs, and example applications, is available at \url{https://github.com/hpcclab/OaaS-IoT}.

\section{Performance Evaluation}
\label{sec:evaluation}
\vspace{-2.5mm}
\subsection{Methodology}




\begin{figure*}[t]
    \centering
    \subfloat[FaaS-based solutions using AWS and Azure services \cite{inventory-managerment-azure}]{
        \includegraphics[width=0.46\linewidth]{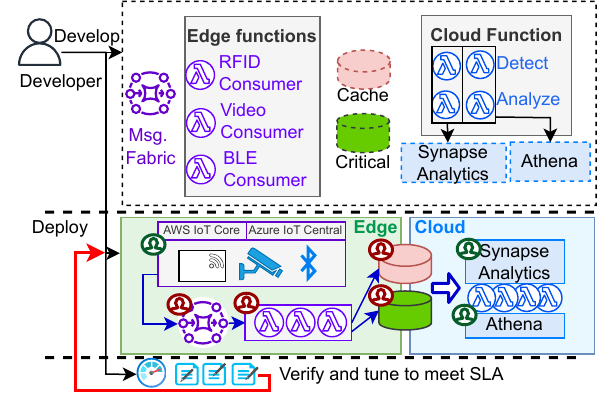}
        \label{fig:evaluation-case-study-faas}
    }
    \hfill    
    \subfloat[Using \name]{
        \includegraphics[width=0.49\linewidth]{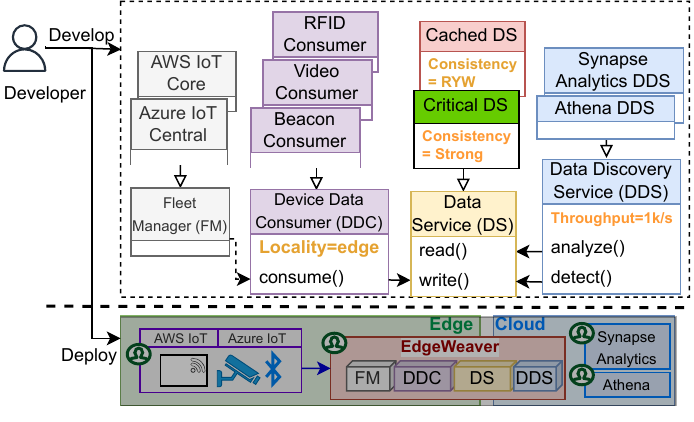}
        \label{fig:evaluation-case-study-oaas-iot}
    }
    \vspace{-1mm}
    \caption{Developing and deploying a real-time inventory management system with FaaS and \name. head and should icons depict system components the developer has to interact during the application life cycle (green: high-level interaction, red: direct configuration). \name needs fewer component interactions and doesn't require the \textbf{verify and tune} loop to meet desired SLAs.
    }
    \label{fig:evaluation-case-study}
    \vspace{-5mm}
\end{figure*}


\textbf{Goals.}
We evaluate \name across realistic settings to assess whether it fulfills its design objectives (\S\ref{sec:realization-object}) and thus, effectively addresses the challenges of IoT application development and deployment across the Edge–Cloud continuum (\S\ref{sec:background}). Specifically, we aim to answer the following key questions:
(i) \textit{\underline{Comprehensiveness and Productivity}:} Does the unified object abstraction and declarative SLA interface provide a high-level view of IoT applications to support diverse QoS requirements and simplify development, ultimately improving developer productivity? (\S\ref{sec:evaluation-results-case-study})
(ii) \textit{\underline{Efficiency for Practice Uses}:} Can \name implement its abstractions and enforcement mechanisms efficiently and at scale, matching or even exceeding the performance of state-of-the-art systems, thus developers enjoy higher productivity without incurring significant trade-offs? (\S\ref{sec:evaluation-results-applicability})
(iii) \textit{\underline{Adaptability for Reliable Execution}:} Can \name dynamically respond to workload and infrastructure changes to preserve QoS with minimal developer effort? (\S\ref{sec:evaluation-results-adaptability})


\noindent\textbf{Experimental Setup.}
We conduct experiments on Chameleon Cloud \cite{chameleon_cloud}, using two clusters to represent the cloud and edge tiers. The \textit{cloud cluster} consists of machines equipped with dual-socket Intel(R) Xeon(R) Platinum 8380 CPUs (240 cores total) and 768 GB of memory. The \textit{edge cluster} uses machines with dual-socket Intel(R) Xeon(R) Gold 6240R CPUs (96 cores total) and 256 GB of memory. To reflect realistic deployment scenarios, we deploy the two clusters in geographically dispersed data centers: TACC (Texas) for the cloud and UC (Illinois) for the edge. The clusters communicate over a standard Internet connection with an average round-trip latency of 33 ms.
Cloud cluster runs a full-fledged Kubernetes distribution using rke2~\cite{rke2} while the edge cluster emulates resource-constrained edge environments with K3d \cite{k3d}, a lightweight Kubernetes distribution, in Docker containers. We configure the cloud Kubernetes with unlimited scaling, while the edge k3d consists of 8 K3d clusters, each with 8 vCPUs and 16 GB of memory. We strictly enforce these limits to accurately emulate edge constraints, utilizing the underlying hardware's full capacity only for scalability stress tests. One machine acts as an IoT gateway, generating synthetic data and issuing invocation requests to services deployed at both the edge and the cloud.
To simulate real-world network disruptions, we use Chaos Mesh \cite{chaosmesh2025} to inject intermittent connectivity faults. We install \name (implementation details in \S\ref{sec:implementation}) alongside other baselines across both cloud and edge clusters.

\subsection{Experimental Results}



\subsubsection{Comprehensiveness and Productivity}
\label{sec:evaluation-results-case-study}
We show how \name improves the application development and deployment productivity via case studies and human evaluations.

\noindent\textbf{Comprehensiveness: Design Comparison.} Figure~\ref{fig:evaluation-case-study} shows how developers could use traditional FaaS and \name to develop and deploy a Real-time Inventory Management (IM) system, a common workflow pattern of modern IoT applications.
The FaaS-based architecture, recommended by Azure IoT~\cite{inventory-managerment-azure} and AWS IoT~\cite{dalba2023reference}, ingests heterogeneous data streams (e.g., RFID tags, beacons, video) from devices registered through Azure IoT Central. Each device type requires a dedicated FaaS function tailored to its protocol (e.g., \texttt{RFIDConsumer} over MQTT, \texttt{VideoConsumer} over TCP). Processed data are stored in a fast \textbf{cache} and later queried by analytics functions (e.g., Analyze) via services like \textbf{Azure Synapse Analytics} for analysis, and results are persisted in a \textbf{critical} database. 

This FaaS-based approach is \textit{complex} and \textit{fragmented}. Developers must manage numerous protocol-specific functions, analytics modules, and data stores, while integrating multiple services (e.g., AWS IoT Core, Azure IoT Central) for cross-platform support. It also lacks native QoS enforcement, forcing a manual verify-and-tune cycle of adjusting function placement, resource allocation, and network settings to meet SLAs.

In contrast, \name \textit{unifies} and \textit{automates} this entire process (Figure~\ref{fig:evaluation-case-study-oaas-iot}). Developers work with high-level object abstractions instead of low-level components. Specialized handlers (e.g., \texttt{RFIDConsumer}) are derived from a reusable Device Data Consumers (DDC) class; cloud-specific integrations are encapsulated within a polymorphic Fleet Manager (FM) class; and data management is streamlined via unified Data Service (DS) and Data Discovery Service (DDS) interfaces. SLAs are attached declaratively (e.g., \texttt{Locality=edge} for latency-sensitive tasks, \texttt{Consistency=strong} for critical data), eliminating the need for manual tuning.

Guided by these declarative policies, the \name runtime automatically handles provisioning, placement, and networking, automatically deploying components like FM and DDC at the edge to meet locality requirements. This full-stack automation removes the manual, error-prone configuration cycle inherent in traditional FaaS systems, enabling faster, more reliable, and maintainable IoT deployments.

\noindent\textbf{Productivity Improvement.}
To quantitatively assess EdgeWeaver’s productivity gains over FaaS approach, we implemented two prototypes of the inventory management application: one using Knative (FaaS-based) and one with \name. Development effort was measured using three metrics: Lines of Code (LoC), Lines of Configuration Code (LoCC), and the number of developer-facing interfaces.

The results show a dramatic reduction in development overhead with \name. The Knative implementation required 666 LoC, while \name achieved equivalent functionality in only 363 LoC (44.5\% reduction). The improvement in configuration effort was even more pronounced: \name needed just 39 LoCC, nearly $10\times$ fewer than Knative (417 LoCC), which involves configuring multiple external services such as RabbitMQ, databases, and triggers. This highlights \namens’s strength in abstracting complex infrastructure management.
Furthermore, as shown in Figure~\ref{fig:evaluation-case-study}, the FaaS-based design forces developers to manage at least seven distinct components, whereas \name consolidates these into only four. Together, these results demonstrate that \namens’s unified, declarative interface and automated orchestration significantly reduce development complexity—making it far more productive, maintainable, and developer-friendly than traditional FaaS-based solutions.

\noindent\textbf{Developer Experience.}
To evaluate how \namens's comprehensiveness and productivity translate into better developer experience, we conducted a human study with 39 college students. While this participant pool may not fully represent seasoned professional developers, it effectively measures the \textit{learnability} and ease of adoption of new paradigms—a critical factor for platform adoption. Participants received a 15-minute tutorial on the FaaS and \name abstractions. The duration was short enough to avoid overwhelming participants, yet sufficiently comprehensive to provide essential understanding. The tutorial was followed by a quiz to assess conceptual comprehension and a programming task to evaluate practical performance.
As shown in Table~\ref{tab:quiz_results} (left), participants scored consistently higher on \name-related quiz questions than on FaaS ones, regardless of prior cloud experience. This indicates that \name is more intuitive and easier to learn. For the programming task (Table~\ref{tab:quiz_results}, right), out of the group who can complete the task in both platforms in time, students completed the assignment 31\% faster using \name while achieving nearly identical code quality (\name: 53.6\% vs FaaS: 52.9\%). The programming task grading was based on a rubric weighing Conceptually Correctness (40\%), Functionally Completeness (40\%), and Code Quality (20\%). These results demonstrate that \name’s unified abstractions and automation not only simplify development but also deliver a measurably better, faster, and more accessible programming experience than traditional FaaS-based approaches.


\begin{table}[ht]
  \centering
  \small
  \setlength{\tabcolsep}{5pt}
  \begin{minipage}[t]{0.56\linewidth}
    \centering

    \hspace*{-8mm}
    \begin{tabular}{|l|c|c|}
      \hline
      \textbf{Cloud fam.} & \textbf{EW} & \textbf{FaaS} \\ 
      \hline
      \hline
      Unfamiliar & 84.6\% & 81.5\% \\
      Basic      & 90.0\% & 80.0\% \\
      Competent  & 92.0\% & 88.0\% \\
      \hline
    \end{tabular}
  \end{minipage}
  \hspace{-0.06\linewidth}
  \begin{minipage}[t]{0.43\linewidth}
    \centering

    \begin{tabular}{|l|c|c|}
      \hline
      \textbf{Metrics} & \textbf{EW} & \textbf{FaaS} \\
      \hline
      \hline
      Time (min.)      & 22.43 & 32.40 \\
      Score (\%)  & 52.85 & 53.55\\
      \hline
    \end{tabular}
  \end{minipage}
  \vspace{-1mm}
  \caption{Human Study results (average): Quiz (left) and Programming (right). (EW=\name)}
  \label{tab:quiz_results}
  \vspace{-3mm}
\end{table}

\vspace{2mm}
\noindent
\colorbox{blue!10}{
\parbox{0.96\linewidth}{
\underline{\textbf{Takeaway}:} \emph{
\name provides a unified, comprehensive abstraction that streamlines development and deployment of IoT applications across the Edge–Cloud continuum.
}}}
\vspace{1mm}

\begin{figure}[t]
    \centering
    \subfloat[Availability = 4 nines]{\includegraphics[width=0.51\linewidth]{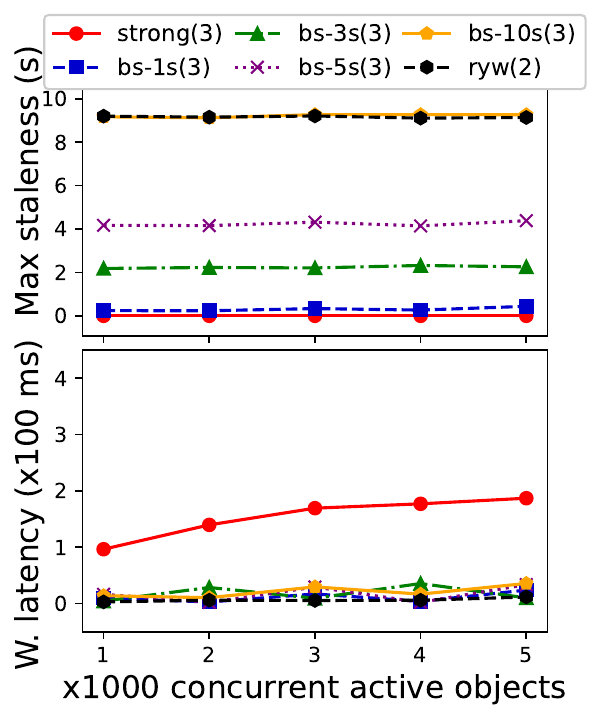}\label{fig:evlt:consistency-4}}
    \hfill
    \subfloat[Availability = 9 nines]{\includegraphics[width=0.49\linewidth]{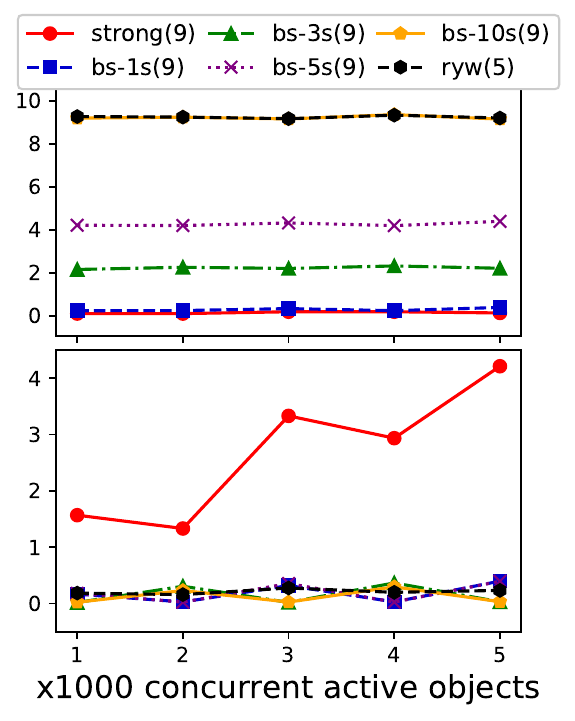}}\label{fig:evlt:consistency-9}
    \vspace{-2mm}
    \caption{Maximum read staleness (top) and average write latency (bottom) under different consistency–availability configurations. Numbers in parentheses show the replica count required to meet both guarantees.}
    \vspace{-4mm}
    \label{fig:evlt:consistency}
\end{figure}

\begin{figure}[t]
    \centering
    \hspace*{-4mm}
    \includegraphics[width=1\linewidth]{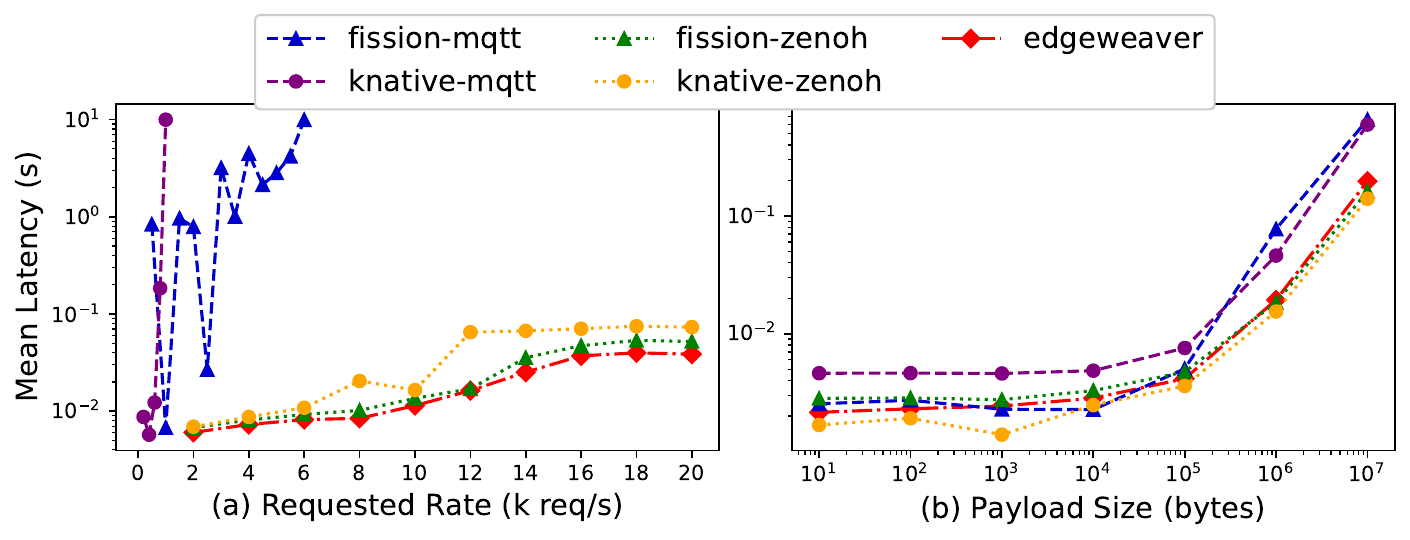}
    \vspace{-2mm}
    \caption{Latency of function invocation with increasing request rate and payload size in various baseline systems.}
    \label{fig:evlt:efficiency}
    \vspace{-5mm}
\end{figure}

\subsubsection{Applicability and Efficiency}
\label{sec:evaluation-results-applicability}
We evaluate applicability of \name by demonstrating its enforcement of diverse SLA combinations through case studies. We assess its efficiency in handling applications with different computational demands, showing its suitability for diverse IoT scenarios.


\noindent\textbf{SLA Enforcement.}
EdgeWeaver's core technical contribution is the automated composition of established distributed algorithms (Raft, CRDTs, Merkle Trees) to enforce declarative SLAs. Here, we evaluate whether this automated composition reliably delivers the specified QoS guarantees across diverse configurations.

We evaluate the ability to enforce various consistency levels: Read-Your-Write (\textit{ryw}), \textit{strong}, and bounded staleness (\textit{bs}) under varying staleness bounds and high availability targets. According to Fig.~\ref{fig:evlt:consistency}, we deploy multiple concurrent \texttt{DataService} objects (from the Inventory case study), each issuing reads and writes. We set availability to 99.99\%, comparable to leading FaaS SLAs (e.g., AWS Lambda’s 99.95\% \cite{aws-lambda-sla}) and Tier-4 datacenters (99.995\% \cite{hpe-cloud-tiers}).

Across all configurations, \name consistently enforces the specified consistency guarantees: Under bounded staleness, observed staleness remains well below set thresholds. Even when stateless is relaxed in RYW, the stateless is consistently below 10 seconds. Notably, under strong consistency, zero staleness is detected, validating \textit{reliable} consistency enforcement of \name.
These guarantees hold at scale: With 1,000 concurrent objects, strong consistency maintains an average write latency of 100 ms, which only increases by $1.87\times$ when scaling to 5,000 objects. RYW and bounded staleness achieve significantly lower latency ($<$ 20 ms), over $9\times$ faster than strong consistency, highlighting promising performance–consistency trade-offs that developers can leverage for various QoS needs.

To test robustness, we increase the SLA target to nine nines (i.e., 99.999999999, five orders of magnitude higher than the current standard. \name continues to satisfy all consistency requirements: bounded staleness remains under 10s, and strong consistency still achieves zero staleness. The write latency for strong consistency increases by at most $2\times$, while weaker models show a negligible impact. Finally, \name achieves these guarantees cost-effectively. At four nines, enforcing availability requires just 2–3 replicas. Even at nine nines, the system needs no more than nine replicas, a $3\times$ cost increase for $10,000\times$ higher reliability.


\noindent\textbf{Implementation Efficiency.}
We evaluate the implementation overhead introduced by \name’s object abstraction and SLA enforcement by comparing it against equivalent FaaS-based implementations. Since \name builds on standard FaaS engines and employs Pub/Sub protocols for communication, we benchmark it using combinations of Knative~\cite{knative} and Fission~\cite{fission} with MQTT~\cite{rabbitmq} and Zenoh~\cite{corsaro2023zenoh}.

Figure~\ref{fig:evlt:efficiency} shows the latency of a lightweight echo function with no workload, representing the intrinsic system overhead of each approach. Although \name introduces additional mechanisms for object abstraction and SLA enforcement, this overhead is negligible. Across different request rates and payload sizes, \name consistently delivers performance on par with or better than the baselines, sometimes even outperforming them (e.g., Knative–MQTT). These results confirm that \namens’s adaptable runtime realizes its object abstraction and declarative SLA enforcement without compromising its performance, providing strong evidence of implementation efficiency in practice.

\begin{figure}[t]
    \vspace{-1mm}
    \centering
    \includegraphics[width=0.9\linewidth]{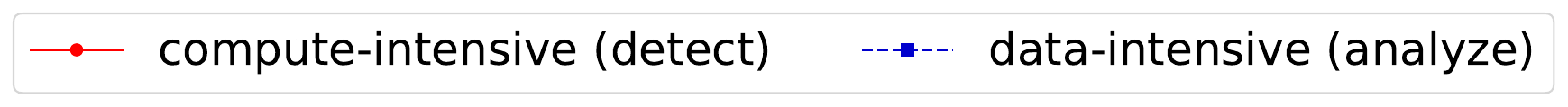}
    \subfloat[Edge-Cloud]{\includegraphics[width=0.28\linewidth]{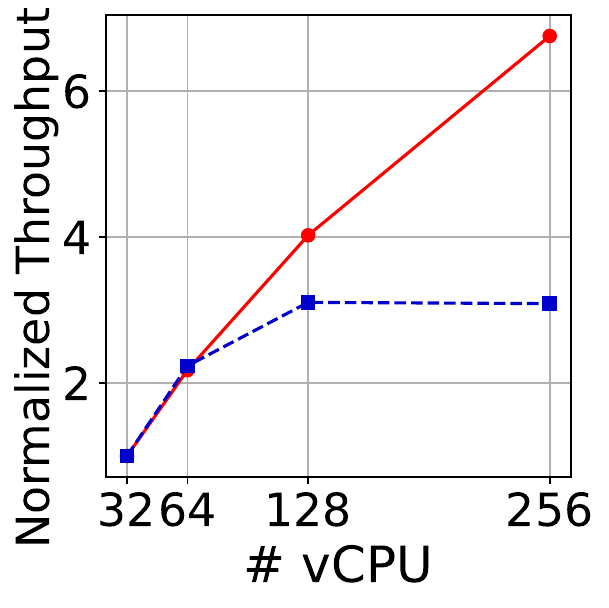}\label{fig:evlt:scale-ec}}
    \hfill
    \subfloat[Cloud]{\includegraphics[width=0.42\linewidth]{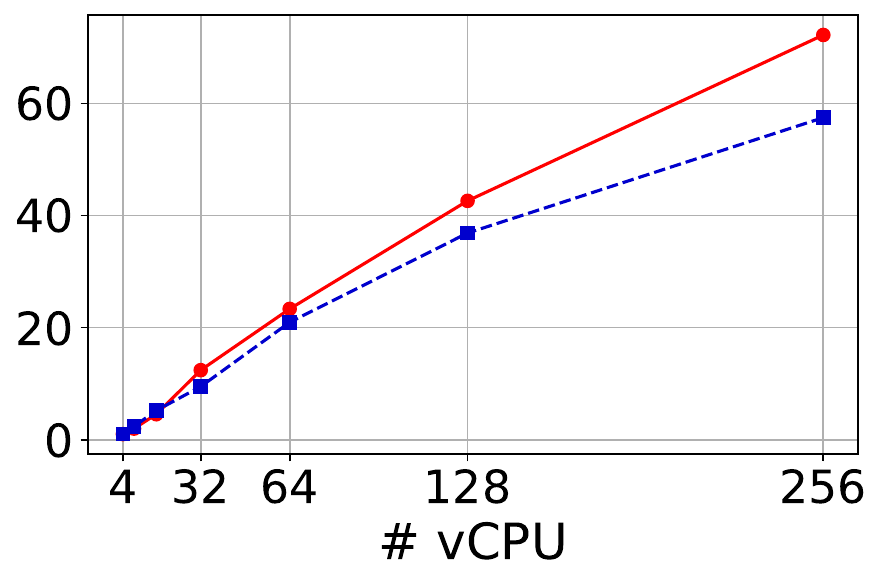}\label{fig:evlt:scale-cloud}} 
    \hfill
    \subfloat[Edge]{\includegraphics[width=0.28\linewidth]{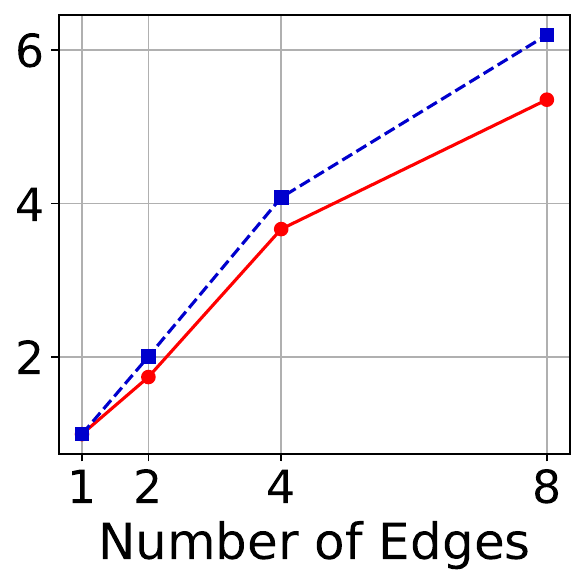}\label{fig:evlt:scale-edge}}
    \vspace{-6pt}
    \caption{Scalability analysis of \name}
    \vspace{-6mm}
    \label{fig:evlt:scalability}
\end{figure}

\noindent\textbf{Scalability.}
We evaluate the scalability of \name by examining whether its implementation efficiency, observed in earlier experiments, holds under workload and resource scaling. To reflect realistic IoT usage, we consider two representative workloads: JSON document processing as a \textit{data-intensive} task and image object detection using the YOLO model as a \textit{compute-intensive} task. Both are implemented as the \textit{analyze} and \textit{detect} functions of the \textit{Data Discovery Service} (DDS) in the inventory management case study (Figure \ref{fig:evaluation-case-study}). For each workload, we deploy multiple object instances across edge and cloud nodes, allowing \name to automatically determine their placement without explicit Locality constraints. Each instance is driven by a dedicated load generator that repeatedly invokes its corresponding function.

Figure~\ref{fig:evlt:scale-ec} shows the throughput as we scale the total number of vCPUs across edge and cloud--from 8 vCPUs at the edge and 24 in the cloud, doubling the capacity incrementally. Throughput is normalized to the lowest allocation. Both workloads show strong scalability: throughput increases nearly linearly up to 128 vCPUs for \textit{analyze} and 256 for \textit{detect}. Beyond those points, performance plateaus due to network saturation between the TACC and UC data centers.
To isolate the network impact, we rerun the experiments independently within each site. Figure~\ref{fig:evlt:scale-cloud} presents the cloud-only results. The \textit{detect} workload, being compute-intensive, scales nearly linearly—achieving a $70\times$ throughput gain from 4 to 256 vCPUs, peaking at 144 invocations/sec. \textit{analyze}, which is more data- and I/O-intensive, scales more moderately, reaching approximately 200,000 invocations/sec at 256 vCPUs.
We observe similar patterns for edge-only (Fig.~\ref{fig:evlt:scale-edge}):  throughput scales proportionally with the number of edges (8 vCPU per edge), confirming \name sustains high throughput and scalability across diverse workloads and deployments.

\vspace{1mm}
\noindent
\colorbox{blue!10}{
\parbox{0.96\linewidth}{
\underline{\textbf{Takeaway}:} \emph{
\name demonstrates strong applicability by providing efficient implementation to reliably enforce diverse SLAs and scale efficiently across Edge–Cloud.
}}}
\vspace{1mm}

\begin{figure} [t]
    \vspace{-1mm}
    \centering
    \includegraphics[width=\linewidth]{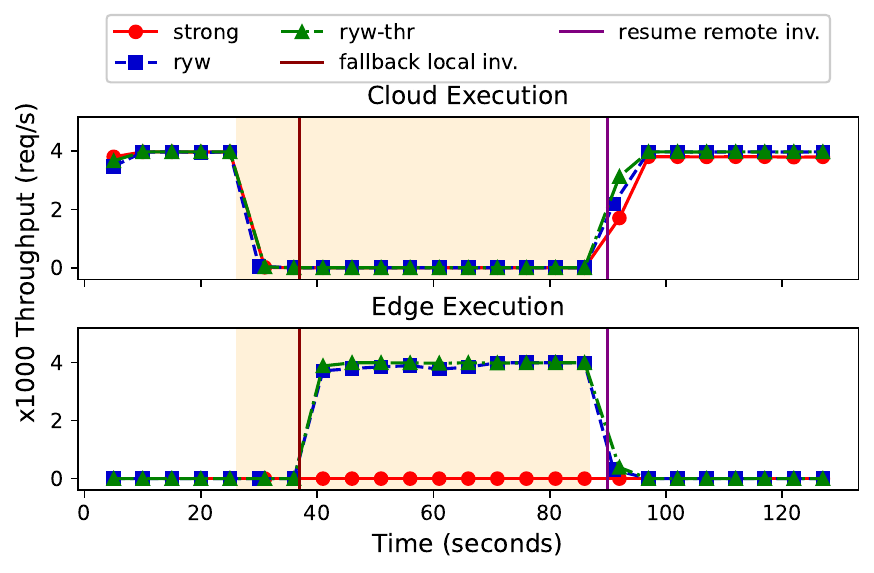}
    \vspace{-7mm}
    \caption{Impact of network partitioning for classes with: RYW, RYW with throughput, and Strong Consistency.}
    \label{fig:evlt:net-partition}
    \vspace{-5mm}
\end{figure}

\subsubsection{Adaptability}
\label{sec:evaluation-results-adaptability}


We evaluate the adaptability of \name by testing its ability to maintain QoS under dynamic network conditions. We deploy three functions with different SLA configurations: (i) \texttt{consume} (from the Device Data Consumer) with \textit{Read-Your-Write} consistency (\texttt{ryw}), (ii) \texttt{write} (from the Data Service) with \textit{RYW + throughput} guarantee (\texttt{ryw-thr}, 4,000 RPS), and (iii) \texttt{detect} (from the Data Discovery Service) with \textit{strong} consistency. We assign high-availability SLAs, prompting \name to place replicas across edge-cloud. We emulate \textit{network partitioning} period between cloud and edge using Chaos Mesh~\cite{chaosmesh2025} (yellow area in Fig.~\ref{fig:evlt:net-partition}), where injected faults disrupt connectivity and trigger \namens's runtime adaptation.

Initially, all functions continuously issue 4,000 RPS write requests to their associated data services. During the partition, \name detects the disruption and redirects invocations to the edge whenever possible.
For the \textit{ryw}, it permits continued execution by relaxing consistency, but since the underlying Logic engine is not inherently prepared for this scenario, its throughput is unstable. In contrast, the \textit{ryw-thr} function benefits from the throughput SLA, maintaining stable performance. For strong consistency, \name enforces quorum strictly; as consensus cannot be achieved across the partition, throughput drops to zero, preserving correctness.
Upon network restoration, \textit{ryw-thr} quickly recovers full throughput, while \textit{ryw} experiences a brief delay due to reactive scaling. The results show \name fine-grained adaptability, enabling  dynamic balancing of QoS desires in response to changes.

\vspace{2mm}
\noindent
\colorbox{blue!10}{
\parbox{0.96\linewidth}{
\underline{\textbf{Takeaway}:} \emph{
SLA-driven deployment enables \name to adapt automatically to dynamic environments to consistently meet application needs.
}}}


\section{Conclusion and Future Works}\label{sec:conclusion}

In this paper, we presented \name, a platform to streamline IoT application development and deployment across the Edge-Cloud continuum. Inspired by OOP principles, \name offers the object abstraction that encapsulates application state, functions (logics), and SLAs, thereby providing a holistic view across the continuum. Moreover, it can transparently handle user-defined consistency and availability trade-offs in the presence of network failure. Importantly, the benefits of \name do not come with any significant overhead to the system.
In the future, we plan to extend \name to support multi-cloud development, allowing objects to run functions across different clouds. 

\section*{ACKNOWLEDGEMENT}
This project is supported by National Science Foundation (NSF) through CNS CAREER Award\# 2419588.

%


\balance
\bibliographystyle{plain}
\bibliography{references}


\end{document}